\documentclass[11pt,twoside]{article} 
\usepackage{acta-info}
\usepackage{euler}
\usepackage{epstopdf}
\usepackage{graphicx}

\newcommand{\vol}{\textbf{2}, 2 (2010) 117--134}   
\newcommand{\CRs}{\CRsubj{%
D.0 
}}   
\newcommand{\amss}{\amssubj{%
68N01 
}}     
\newcommand{\keyw}{\keywords{%
emulation, plug-ins, communication
}}

\begin{document}

\title{%
Communication model of emuStudio emulation platform 
}
\maketitle

\oneauthor{%
\href{http://hron.fei.tuke.sk/~jakubco}{Peter Jakub\v{c}o}
}{%
\href{http://dci.fei.tuke.sk/}{Department of Computers and Informatics}\\
\href{http://www.fei.tuke.sk/en/faculty/}{Faculty of Electrical Engineering and Informatics}\\
\href{http://www.tuke.sk}{Technical University of Ko\v{s}ice} 
}{%
\href{mailto:peter.jakubco@tuke.sk}{peter.jakubco@tuke.sk}
}

\twoauthors{%
\href{http://hornad.fei.tuke.sk/kpi/person/simonak/kpicard.php}{Slavom\'ir \v{S}imo\v{n}\'ak} 
}{%
\href{http://dci.fei.tuke.sk/}{Department of Computers and Informatics}\\
\href{http://www.fei.tuke.sk/en/faculty/}{Faculty of Electrical Engineering and Informatics}\\
\href{http://www.tuke.sk}{Technical University of Ko\v{s}ice} 
}{%
\href{mailto:slavomir.simonak@tuke.sk}{slavomir.simonak@tuke.sk} 
}{%
\href{http://hornad.fei.tuke.sk/kpi/person/adam/kpicard.php}{Norbert \'Ad\'am}
}{%
\href{http://dci.fei.tuke.sk/}{Department of Computers and Informatics}\\
\href{http://www.fei.tuke.sk/en/faculty/}{Faculty of Electrical Engineering and Informatics}\\
\href{http://www.tuke.sk}{Technical University of Ko\v{s}ice} 
}{%
\href{mailto:norbert.adam@tuke.sk}{norbert.adam@tuke.sk} 
}

\short{%
P. Jakub\v{c}o, S. \v{S}imo\v{n}\'ak, N. \'Ad\'am
}{%
Communication model of emuStudio emulation platform
}

\begin{abstract}
Within the paper a description of communication model of plug-in based emuStudio emulation platform is given. The platform mentioned above allows the emulation of whole computer systems, configurable to the level of its components, represented by the plug-in modules of the platform. Development tasks still are in progress at the home institution of the authors. Currently the platform is exploited for teaching purposes within subjects aimed at machine-oriented languages and computer architectures. Versatility of the platform, given by its plug-in based architecture is a big advantage, when used as a teaching support tool. The paper briefly describes the emuStudio platform at its introductory part and then the mechanisms of inter-module communication are described. 
\end{abstract}


\section{Introduction}
Emulation is an imitation of internal structure of the system, by which we simulate its behavior or functionality. An emulator can be implemented in a form of software, which emulates computer hardware – its architecture and functionality as well. 

Development of a fully-fledged emulator is connected with many areas of computer science, like a theory of compilers (needed mainly at instructions decoding in emulated processor), theory of emulation (includes different algorithms of emulation, methods of abstraction of real hardware into its software model), programming languages and programming techniques (required for performance improvements) and obviously detailed knowledge of emulated hardware.

The goal of our effort connected with the emuStudio was an emulation platform able to emulate different computers, which are similar by their structure. Such a tool was intended to be a valuable utility supporting the teaching process in areas like machine-oriented languages and computer architectures, so simplicity and configurability were the properties also considered.  
Thanks to the universal model of plug-in communication, we were able to create emulators of different computers like a real MITS Altair8800~\cite{ALTAIR} in two variations, or an abstract RAM machine~\cite{RAM}, and others. More information on this topic can be found in~\cite{CPUEMU,EMUSTUDIO}.

Not too much universal emulators are available nowadays. Partial success with standardizing emulated hardware was achieved within a project M.A.M.E. \cite{MAME}, which is oriented toward preserving historical games for video game consoles. 
In 2009 a medium scale research project co-financed by the European Union's Seventh Framework Programme started with the aim to develop an Emulation Access Platform (KEEP)~\cite{KEEP}. As far as we know, the ability provided by the emuStudio platform to choose the configuration of emulated system by the user dynamically is unique.

\section{Architectures for emulation}

Computer architecture is a system characteristic, which unifies the function and the structure of its components~\cite{APS}. Most widely known architectures are Harvard architecture~\cite{DCP} and Princeton architecture (also known as von Neumann architecture)~\cite{EDVAC}, which is the core of most modern computers.  

Versatility of the platform is oriented towards a liberty in choosing the configuration, rather than architecture of emulated computer. Configuration choice is given by the selection of components (plug-in modules) and their connections (a way of communication). 

\begin{figure}[!ht]
\centering
\includegraphics[width=0.5\textwidth]{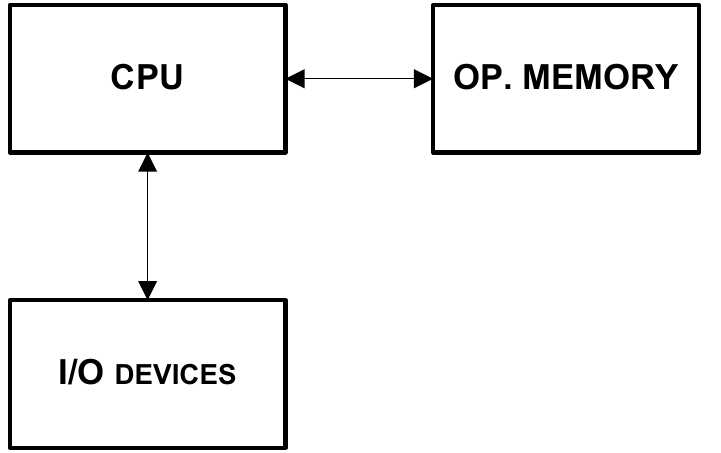}
\caption{Computer architecture of von Neumann type}\label{FIG:VONNEUMANN}
\end{figure}

As a basic architecture for emulation, the von Neumann architecture type was chosen (Figure~\ref{FIG:VONNEUMANN}). Communication model (methods, protocol) thus was adapted for this type of architecture. 

The core component of the architecture is a processor (CPU), which executes instructions, communicates with main memory, from which it fetches instructions to execute. Main memory, except the instructions mentioned, also stores data. CPU also communicates with peripheral devices. An extension of the emulator, compared to a basic von Neumann concept is that peripheral devices are allowed to communicate each other without the interaction of CPU and also with main memory.

The selection of computer configuration to emulate is left to the user. Required configuration thus can be composed by picking and connecting available components by the user. From the configuration composed a platform creates an instance, when the one is selected at the startup. By this selection, the virtual architecture arises, ready for emulation.

\begin{figure}[!ht]
\centering
\includegraphics[width=0.5\textwidth]{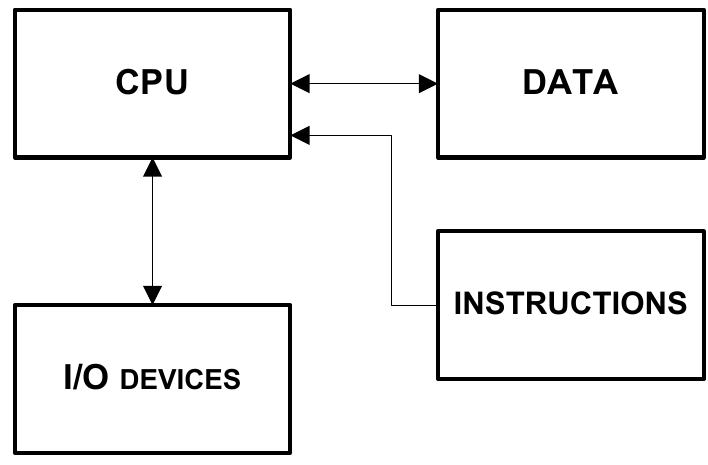}
\caption{Computer architecture of Harvard type}\label{FIG:HARVARD}
\end{figure}

The fact that the communication model is adapted for configurations of von Neumann type, does not guarantee any support for creating architectures of different type, but also it does not exclude it. As an example can serve the implementation of RAM machine emulator at our platform, which in fact uses the architecture of Harvard type.

Main difference between the two architectures mentioned is, that computers with Harvard architecture (Figure~\ref{FIG:HARVARD}) use two types of memory – first for storing instructions, second for storing data.

\section{The structure of the platform}

Basic components of a computer from any of architecture types mentioned above can be subdivided into three types:

\begin{itemize}\addtolength{\itemsep}{-0.5\baselineskip}
\item Processor (CPU),
\item Operating memory,
\item Input/output peripheral devices.
\end{itemize}

Particular components of real computers are interconnected by communication elements, like buses. Except that, components make use a number of support circuits, performing auxiliary functions. 

When abstractions of real computers are considered (what emulators surely are), such elements usually are omitted, because they are not essential for proper functionality at given level of abstraction. Although emulation of buses would take us closer to real computer structure, the communication of components emulated would be non-effective (bus as a useless component in between other components) and can introduce possible difficulties (e.g. when data of greater size than the bus width are to be transferred). That's why buses are not used and components emulated use different way of communication (e.g. direct call of components' operations). When identifying, what would be and what would be not implemented in an emulator, it is necessary to take into account the emulator properties required. When, for example, the communication path tracking is required, then emulation of buses and communication elements is necessary. 

\begin{figure}[!ht]
\centering
\includegraphics{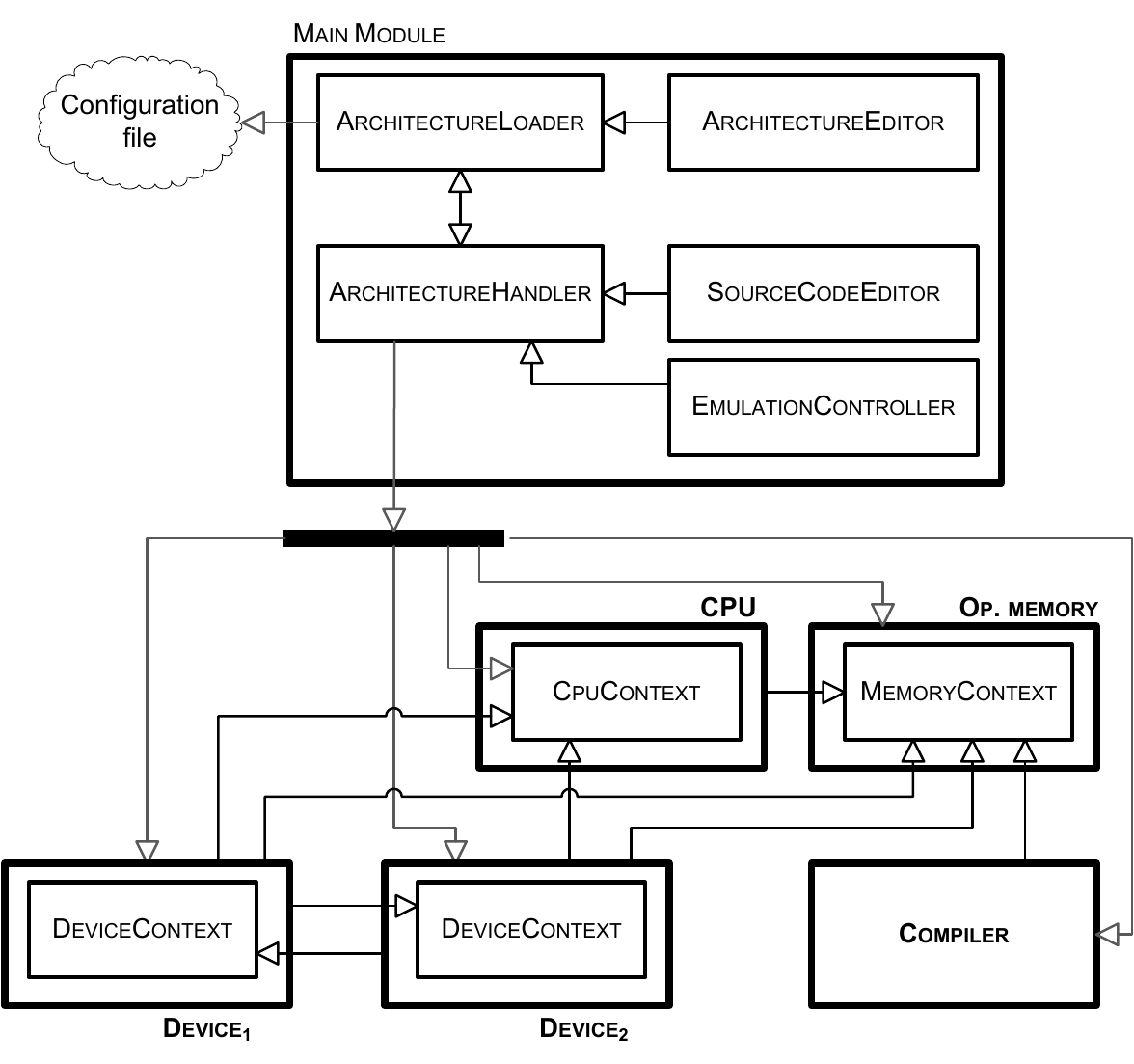}
\caption{The structure of emuStudio platform}\label{FIG:EMUSTUDIO}
\end{figure}

The structure of emuStudio platform is depicted in Figure~\ref{FIG:EMUSTUDIO}. Within the scheme, the arrow direction has the 
following meaning: let's have two objects $O_1$ and $O_2$ from the scheme. When the arrow points from $O_1$ to $O_2$, ($O_1 \rightarrow O_2$), then object $O_1$ is allowed to call operations of object $O_2$, but not in the opposite direction ($O_2$ only can return the result of an operation). According to the scheme, four types of plug-in modules exist in the platform:

\begin{itemize}\addtolength{\itemsep}{-0.5\baselineskip}
\item Processors (CPU),
\item Memories,
\item Input/output peripheral devices,
\item Compilers.
\end{itemize}

The emuStudio platform is implemented using the Java programming language, so one of advantages is the portability at a machine-code (byte-code) level. On the other side, Java programs itself are emulated too, by the Java virtual machine, so the emulator performance is decreased by this choice. 

\subsection{Context}
As it can be seen in Figure~\ref{FIG:EMUSTUDIO}, plug-in modules, except the compiler, contain a special component called context. Lines connecting modules, start from the edge of the module (e.g. \textsc{Device2}), but point to the context of another module (\textsc{CpuContext}). It means that plug-in module, which requests another one, has no full access to this module. It can access the context of this module only. 

Reasons for creating a component context are three: \textit{safety}, \textit{functionality encapsulation} and allowing for \textit{non-standard functionality} of plug-in modules. 

Plug-in module within the emuStudio platform is safe, if its functionality cannot be abused. We mean by this the functionality of module itself (e.g. unwanted change of internal parameters of the module, defined by the user), but also the functionality of the platform as a whole (e.g. controlling the emulation, terminating the application, or unwanted "dynamic" changes in virtual configuration).

For that reason the main module is the only one that has an access to all plug-in operations, and plug-ins consider the main module to be trusted.

Besides, communication model and API (\textit{Application programming interface}) of plug-ins are open and free, so practically the plug-ins can be designed by anyone, by what the credibility of plug-in decreases. The safe functionality therefore should be separated, what has implied to context creation.

On the other hand, it is also not good idea if the plug-ins allow to use a functionality by another plug-ins that the other side doesn't need. The transparency fades out and there again arises the risk of improper use of the operations. The encapsulation principle used in Object oriented programming paradigm therefore claims to hide such operations, what is solved by the use of the context, too.

It is enough if the context will define only the standard functionality (in the form of communication operations) for plug-ins of equal types. However a situation can arise there, wherein this communication standard doesn't have to universally cover all the requirements of each concrete plug-in.

The context is therefore an ideal environment, where such non-standard functionality can be implemented. The fact that the context is realized inside a plug-in, enables to add operations that are not included in the standard context, into the plug-in implementation. Plug-ins that use the non-standard functionality have to know the form of a non-standard context -– otherwise they cannot use the non-standard functionality.

\newpage
\subsection{Main module}
The core of the platform is the main module. It consists of several components:

\begin{description}
\item[ArchitectureLoader] –- the configuration manager. It manages configuration file (stores and loads defined computer configurations), and creates an instance of virtual configuration (through \textsc{ArchitectureHandler} component).

\item[ArchitectureEditor] -– configuration editor. The user uses this component to choose and connect components of defined computer architecture. This selection and connection is realized in a visual way – by drawing of abstract schemas. The component allows creating, editing and deleting the abstract schemas, and it cooperates with \textsc{ArchitectureLoader} component.

\item[ArchitectureHandler] -– the virtual architecture instance manager. It offers plug-ins instances to other components (other plug-ins) and implements an interface for storing/loading of plug-ins' settings.

\item[SourceCodeEditor] -– the source code editor. It allows creating and editing the source code for chosen compiler, it supports syntax highlighting, rows labeling and directly communicates with the compiler (through \textsc{ArchitectureHandler} component).

\item[EmulationController] -– the emulation manager. It controls the whole emulation process, and it stands in the middle of the interaction between virtual architecture and the user.
\end{description}

\subsection{Compiler}
The compiler plug-in represents a translator of source code into a machine code for concrete CPU. The structure of compiler's language is not limited at all, therefore the language doesn't have to be an assembler.

The compiler is chosen by the user in the configuration design process of emulated computer (such as other components are). The logical is a choice of compiler that compiles the source code into machine code for a processor chosen. 

The output of the compiler should be a file with a machine code and optionally the output is redirected into operating memory, too. It depends on a concrete compiler, how the output will be realized.

\subsection{CPU}
Central processing unit (CPU) is a component of digital computer that interprets instructions of computer program and process data. CPU provides fundamental computer property of programmability, and it is one of the most significant components found in computers of each era, together with operating memory and I/O devices.

CPU plug-in represents a virtual processor. It is a base for whole emulation, because the control of emulation run in the main module actually means the control of processor run. The main activity of a CPU is the instruction execution. These instructions can be emulated by arbitrary emulation technique~\cite{CPUEMU} (depending on implementation of a concrete plug-in).

The plug-in contains a special component called \textit{processor context}, operations of what are specific for concrete CPU (besides the standard operations, there can be specified more operations by the programmer). Devices that need to have an access to the CPU get only its context available. Therefore the context for devices represents a "sandbox" that prohibits interfering with sensible settings and the control of processor's run. If such a device has to be connected with CPU, the context should contain operations that allow device connections.

\subsection{Operating memory}
The operating memory (OP) represents a virtual main store (storage space for data and instructions). Generally an OP consists from cells, where format, type, size and value of cells are not closely defined. Cells are placed sequentially, therefore it is possible to determine unique location of any cell in the memory (the location is called an \textit{address}).

OP contains a component called \textit{memory context} that besides the standard operations (reading from and writing to memory cells) can include specific operations (e.g. support of segmentation, paging and other techniques), too, of a concrete plug-in. 

Devices (and a compiler) that need to have an access to OP (e.g. devices that use direct access into memory), get only this memory context. Devices get the context in a virtual architecture initialization process, and compiler (when needs to write compiled code directly into memory) when calling the \textit{compile} method. Therefore operations in the context have to be safe (from usability's point of view) for other plug-ins.

\newpage

\subsection{Peripheral devices}
Peripheral devices are virtual devices that emulate functionality of real devices. Generally the purpose of the devices is not closely defined, nor standardized, so plug-ins do not really need represent real devices.

The main idea of device communication is that all information within the communication process go into the device though its input(s) and go out from the device as one or more outputs. The devices then can be input, output, or input/output. 

In some detail (that detail is not limited), the devices can work independently (and reacts to events of connected plug-ins), eventually interacts with the user. Devices can communicate with CPU, and/or OP, and/or other devices.

The communication model supports hierarchical device connections (the devices are therefore enabled to communicate to each other without the CPU attention/support). 

Every device (following the Figure~\ref{FIG:EMUSTUDIO}) contains one or more components, called \textit{device context}. The device context can be extended by a concrete plug-in with non-standard operations. Other devices that need an access to this device get one or more contexts of the device (that mechanism ensures that one device can be connected to more devices). The operations available in the context have to be safe (from usability's point of view) for other plug-ins.

\section{Communication realization in emuStudio\\ platform}
As could be seen, plug-in contexts solve some concrete problems of communication module. In this section, communication model will be described in more detail, and a way how the communication is realized between the main module and plug-ins. The communication model represents a collection of standardized methods, and by calling of them individual sides will communicate with each other.

\begin{figure}[!ht]
\centering
\includegraphics[width=0.6\textwidth]{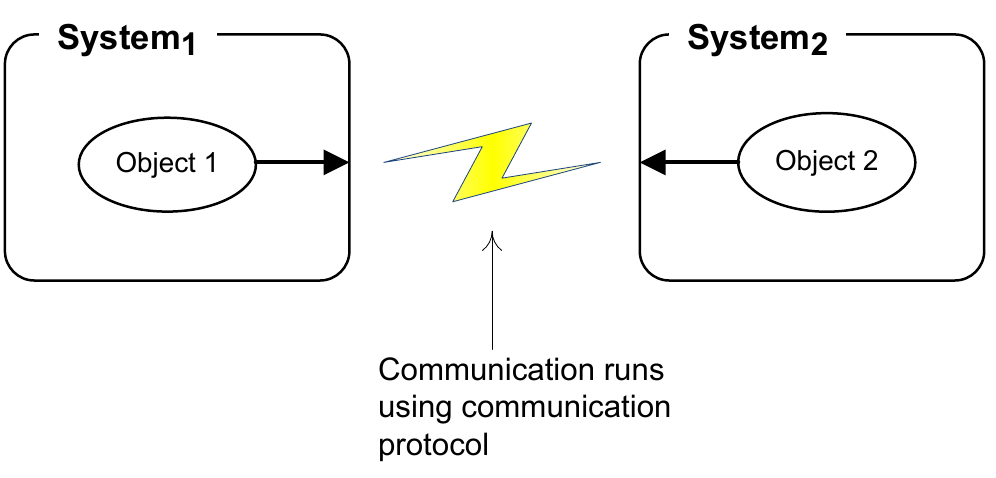}
\caption{Communication realization between two independent and separated objects}\label{FIG:SEP}
\end{figure}

Let's consider two objects that want to communicate with each other. In the case when communication sides are independent and separated systems (one side cannot directly call the other side), it is necessary to design a communication protocol and realization mechanism of the communication (e.g. medium) that are somewhat "bridge above the communication gap" (e.g. network) between the objects (Figure~\ref{FIG:SEP}).

The other case arises if the first object can directly access to the second object. The communication in this case will run directly, i.e. objects will directly call operations of the other objects. If the objects are separated and independent, two questions can arise.

\begin{figure}[!ht]
\centering
\includegraphics[width=0.9\textwidth]{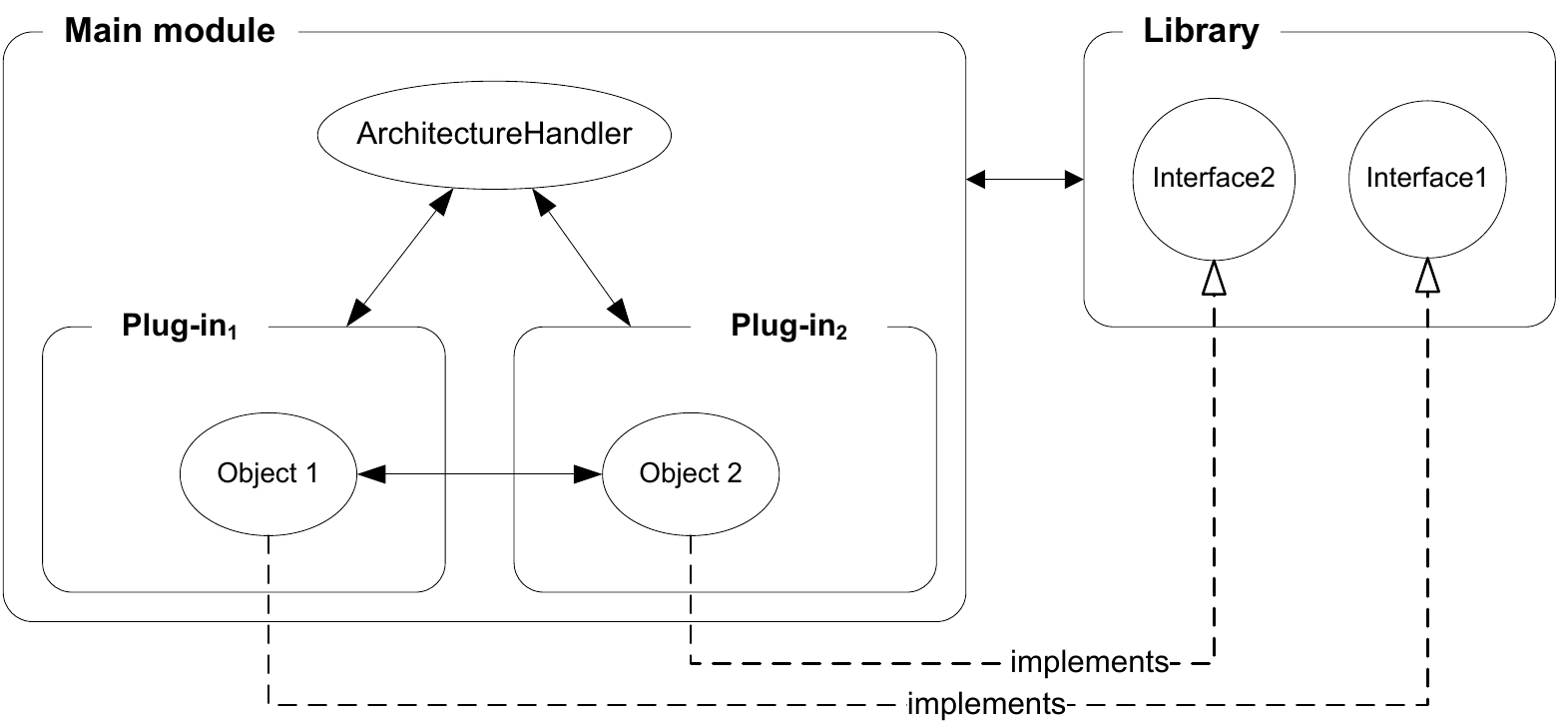}
\caption{Communication realization in emuStudio platform}\label{FIG:TOG}
\end{figure}

At first, how the objects get access to other objects? The solution is to use another, third system, that will cover both subsystems (where the objects reside), or if you like will have direct access to both communicating objects, and the system will \textit{provide} these objects each to another. Just in that way the communication in emuStudio platform works (Figure~\ref{FIG:TOG}).

The second question is, how can the objects communicate together, that truly are of the same type, but belong to different systems with different implementations? 

The solution is to create a standard model of operations that all the objects of given type will implement and it will be well-known to all objects. This model has to unify both the syntax and semantics of communication operations.

The main module represents a system of higher level, covering subsystems – plug-ins. The objects of plug-ins the main module will get in virtual architecture instance creation process.

Communication operations are well-known both by the main module and by plug-ins. This is ensured by the fact that \textit{prototypes} of the operations lies in external library, where the access is granted to both the main module and plug-ins. The operations are \textit{ordered} according to the plug-in \textit{type} into interfaces (an \textit{interface} is defined as a structure containing a list of operations without their implementation). Each plug-in type has its own set of interfaces that corresponding plug-in has to implement.

\subsection{External library structure}
\begin{figure}[!ht]
\centering
\includegraphics[width=0.7\textwidth]{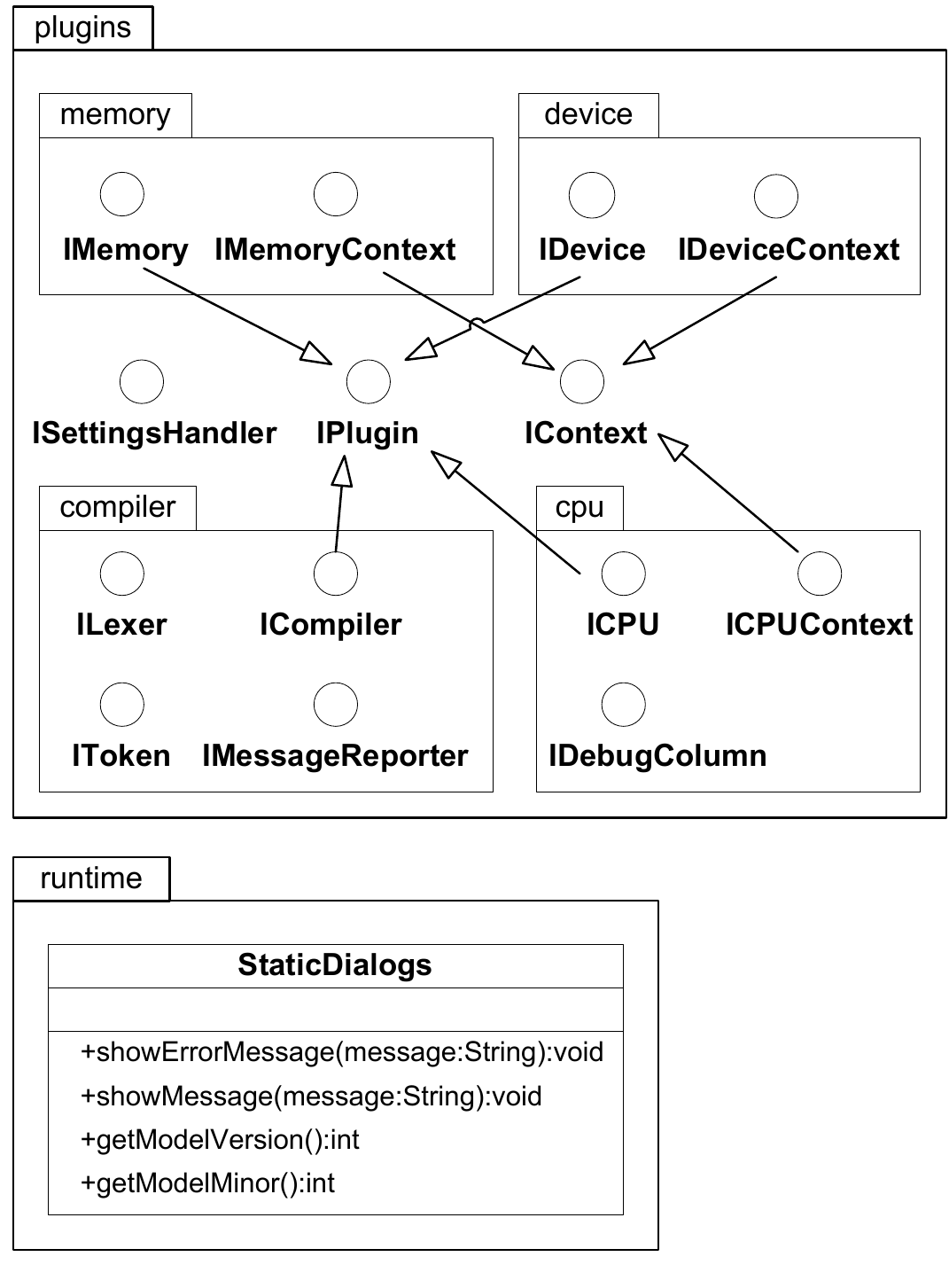}
\caption{Library structure}\label{FIG:LIB}
\end{figure}

Figure~\ref{FIG:LIB} shows the structure of the external library, and contains all the prototypes (interfaces) for plug-ins.

Besides the packages and interfaces intended for the plug-ins to use, it contains a class called \verb+runtime.StaticDialogs+, too. The class contains static methods that should disburden the plug-ins from common, fundamental and very often used methods implementation.

\subsection{Standard operations -- Compiler}
Every compiler generally consists of these parts:

\begin{itemize}\addtolength{\itemsep}{-0.5\baselineskip}
\item Lexical analyzer,
\item Syntactic analyzer (parser) that builds abstract syntactic tree,
\item Semantic analyzer that verifies types usage and other semantic information,
\item Code generator that generates a machine code using abstract syntactic tree.
\end{itemize}

Only some of these parts are important for interaction with the main module and plug-ins. Lexical and syntactic analyzer need to have the access to the source code itself. Semantic analyzer can work with knowledge gained from the phase of syntactic analysis (abstract syntactic tree), it means that semantic analyzer won't be in direct interaction with main module or other plug-ins. Therefore it is possible to skip all considerations of assigning it into a communication model.

Machine code generator can have an access to operating memory, too – if the user asks to redirect the compiler output into operating memory. On the other hand, the main module needs to have an access to lexical analyzer, in order to make possible to use syntax highlighting in source code editor. Finally, the main module needs to call the compile operation itself.

In  Table~\ref{TAB:COMPILER} basic standard operations are described that are important from the communication point of view.

\begin{table}[!ht]
\centering
\begin{tabular}{|l|l|}
\hline
\textbf{Operation} & \textbf{Description} \\
\hline
\verb+Compile+ & Source code compiling \\
\hline
\verb+GetLexer+ & Gets an lexical analyzer object \\
\hline
\verb+GetStartAddress+ & Gets absolute starting address of compiled program.\\
                       & The address can be later used as starting address for\\
                       & the program counter after the CPU \textsc{Reset} signal.\\
\hline
\end{tabular}
\caption{Standard compiler operations}\label{TAB:COMPILER}
\end{table}

\subsection{Standard CPU operations}
Processor, or if you like the CPU, is a core of the architecture. It realizes the execution of the whole emulation, because its main activity is instruction execution. It also interacts with peripheral devices and with operating memory. Communication model does not limit the usage of the emulation technique for the processor emulation.

The CPU plug-in in the emuStudio platform besides the emulation itself, it has to co-operate with the user – by the interaction using debugger and status windows (however operations related to the interaction will not be described here). In the status window the CPU should show the values of its registers, flags, actual CPU's running state and eventually other characteristics. The plug-in includes complete status window implementation so with different CPU the content of the status window will change accordingly. 

Generally each CPU plug-in consists of following parts:
\begin{itemize}\addtolength{\itemsep}{-0.5\baselineskip}
\item The processor emulation implementation,
\item Processor context that extends its functionality,
\item Instruction disassembler,
\item Status window GUI.
\end{itemize}

The CPU plug-in design demands the programmer to know the hardware that he is going to implement and to "answer the questions correctly" – when the interface methods implementation are considered.

\begin{figure}[!ht]
\centering
\includegraphics[width=0.4\textwidth]{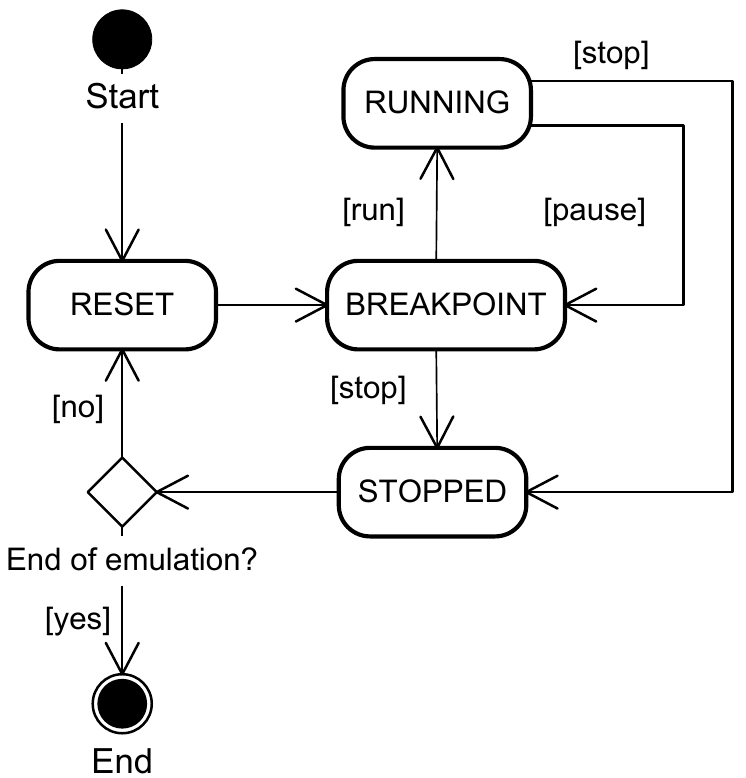}
\caption{Processor work-flow cycle}\label{FIG:WORK}
\end{figure}

The work-flow cycle of each processor plug-in for the emuStudio platform is shown in Figure~\ref{FIG:WORK}. As it can be seen from the figure, the processor can be found in one of four states. The \textsc{Reset} state is that state in which the processor re-initializes itself and immediately after finishing that it sets itself to the \textsc{Breakpoint} state.

For the processor execution only the three states are meaningful:
\begin{itemize}\addtolength{\itemsep}{-0.5\baselineskip}
\item \textsc{Breakpoint} –- the processor is temporally inactive (paused)
\item \textsc{Running} –- the processor is running (executing instructions)
\item \textsc{Stopped} –- the processor is stopped (waits for \textsc{Reset} signal)
\end{itemize}

The Table~\ref{TAB:CPU} describes basic operations to control the processor execution in the communication model. These operations tell how the CPU behavior can be influenced. However all of the operations are not supported in the real CPU's world and by contrast there definitely exist some CPU control operations that are not covered by the communication model. But mostly such operations are not common for all CPUs; therefore their support is optional within the scope of CPU context.

\begin{table}[!ht]
\centering
\begin{tabular}{|l|l|}
\hline
\textbf{Operation} & \textbf{Description} \\
\hline
\verb+Reset+ & Re-initialization. The operation sets the CPU into a state in\\
             & which it should be right after CPU launch (\textsc{Reset} in\\
             & Figure~\ref{FIG:WORK}).\\
\hline
\verb+Step+  & Emulation step. The CPU executes one instruction and then\\
             & returns to the \textsc{Breakpoint} state.\\
\hline
\verb+Stop+  & The operation stops running or paused emulation. The CPU\\
             & leads itself into a state in which it is not able to execute\\
             & instructions anymore, till its reset (\textsc{Stopped} in Figure~\ref{FIG:WORK}).\\
\hline
\verb+Pause+ & Block/pause running emulation. The CPU leads itself to a\\
             & state in which it stops to execute instructions, but its state\\
             & (register values, flags and other settings) is unchanged after\\
             & the last executed instruction (\textsc{Breakpoint} in Figure~\ref{FIG:WORK}).\\
             & From this state the CPU can be launched again.\\
\hline
\verb+Execute+ & The operation launches paused/blocked emulation. The CPU\\
               & leads itself to a state in which permanently executes instructions\\
               & (\textsc{Running} in Figure~\ref{FIG:WORK}). The stop of the CPU in this state can\\
               & be activated by the user, otherwise the CPU stops spontaneously\\
               & (e.g. after the execution of halt instruction).\\
\hline
\end{tabular}\caption{Some of the standard CPU operations}\label{TAB:CPU}
\end{table}

\subsection{Standard operations -- Operating memory}
Operating memory (OP) is not a computer component that directly affects other computer components. It means that the memory is not ``demanding'' for services – it is not acting like a communication initiator with the CPU, nor with the other devices (according to von Neumann conception). This fact is covered by the communication model – all connections with the OP are one-directional, and the OP is always plugged \textit{into} the device (or into a processor), and not in the other way. It means that the device (or processor) can use services of OP, but the OP cannot use services of the device – the OP doesn't need to have an access to any plug-in.

The programmers can use the memory context also for the implementation of methods that allow attaching devices into operating memory. Such type of connection can be useful, if a device needs to be informed of the OP changed status (e.g. DMA technology).

Each OP implementation has to include a graphical user interface (GUI) – so each memory should provide a graphical view to its content for a user (the content is represented by the values of its cells) and eventually to provide another manipulation with memory (e.g. address or value searching, memory content export into a file, etc.). Executed processor instructions description and the graphical view of memory content are basic interaction resources that the user has a contact with.

Summarizing previous paragraphs there can be named all components that each OP plug-in must contain:

\begin{itemize}\addtolength{\itemsep}{-0.5\baselineskip}
\item Memory context,
\item Implementation of main interface –- the memory functionality itself,
\item Graphical user interface (GUI).
\end{itemize}

Basic operations that have to be implemented in each operating memory are described in Table~\ref{TAB:MEM}.

\begin{table}[!ht]
\begin{tabular}{|l|l|}
\hline
\textbf{Operation} & \textbf{Description} \\
\hline
\verb+Read+ & Reading from operating memory -– either one or more cells at\\
            & once starting from given address.\\
\hline
\verb+Write+ & Writing into operating memory -– either one or more cells at\\
             & once starting from given address.\\
\hline
\verb+ShowGUI+ & The operation shows graphical user interface (GUI) of\\
               & memory content.\\
\hline
\end{tabular}\caption{Some of the operating memory standard operations}\label{TAB:MEM}
\end{table}

\subsection{Standard operations -– Peripheral devices}
There are known input, output and input-output devices. Their category can be identified easily according to a way how they are connected with other components of the configuration and to the direction of the connection (direction of data flow).

It is possible to implement virtual devices that communicate with real devices, but also fictive and abstract devices can be implemented. The device can interact with the user through its own graphical interface (GUI). Not all devices have to have GUI, but on the other hand there are such devices that their input and/or output are realized using the user interaction (e.g. terminals, displays). The devices can communicate with CPU, OP and with other devices, too.

A single device can be connected multiple times with other components. For this reason the devices can have several contexts, with possible different implementations. For example a serial card can have several physical ports, into which it is possible to plug in various devices (into each port can be plugged a single device, and each port is represented by a single context).

Communication model solves the following problems:

\begin{itemize}\addtolength{\itemsep}{-0.5\baselineskip}
\item How to connect devices to each other,
\item How to realize input/output.
\end{itemize}

The basic idea of interconnection of two devices in the meaning of implementation is their contexts exchange with each other. All input/output operations that the devices will use for the communication resides in the context. In such a way the bidirectional connection is realized. Each device contains an operation intended for attaching of another device (op. \verb+attachDevice+), that as a parameter takes the context of connecting device. This connection operation does not reside in the context, in order to ensure that the plug-ins couldn't change the structure of the virtual architecture. The connection job itself does the main module that performs the interconnection only in the virtual architecture creation process.

The input and output operations (\verb+in+ and \verb+out+) reside in the device context, because by calling them the communication is performed. Transferred data type in these operations is not specified in the communication model, but it is defined by the plug-ins. The java \verb+Object+s are therefore transferred (they are returned by the \verb+in+ method and the \verb+out+ method uses it as a parameter).

\section{Conclusions}
As far as we know, the emuStudio platform is the first attempt of the implementation of both the universal and interactive emulation platform with the emulated components realized via plug-ins. In the present time the platform is used as a teaching support tool for chosen subjects at the Department of Computers and Informatics, Faculty of Electrical Engineering and Informatics, Technical University of Ko\v{s}ice, Slovakia, in its still expanding form for more than two years.

The versatility and configurability allows creating plug-ins of various levels of quality and purpose -– they can be intended for pedagogic or even for scientific purposes -– e.g. the implementation of plug-ins that emulate the real hardware with the support of measurement of various characteristics, or as one of the phases of design of new hardware or for its testing, etc.

For ensuring the platform's versatility it is important to stabilize the requirements, to standardize components and mainly to design a way of communication in the form of communication protocol, language or other mechanism. 

The paper describes the mechanism of communication used in the emuStudio platform at the basic level. The communication mechanism still is not in its final form. Till the present time the 8-bit architectures (MITS Altair8800 and its modification) and two abstract machines (Random Access Machine, BrainDuck –- our own architecture) are implemented only.

We believe that in the future the platform will be enhanced and the communication model finished and formally verified. There still is a free space for expanding the platform by adding new emulated computer architectures.




\bigskip
\rightline{\emph{Received: March 12, 2010 {\tiny \raisebox{2pt}{$\bullet$\!}} Revised:	June 25, 2010}}     

\end{document}